\documentstyle{article}
\begin{document}
\title{QRPA and its extensions in a solvable model\\~\\
QRPA y sus extensiones en un modelo soluble exactamente}

\author{Jorge G. Hirsch$^1$\thanks{e-mail: hirsch@fis.cinvestav.mx},
	Peter O. Hess$^2$\thanks{e-mail: hess@roxanne.nuclecu,unam.mx}
 and
	Osvaldo Civitarese$^3$\thanks{e-mail:
civitare@venus.fisica.unlp.edu.ar}\\
{\small\it $^1$Departamento de F\'{\i}sica, Centro de Investigaci\'on
y de Estudios Avanzados del IPN,}\\
{\small\it Apdo. Postal 14-740 M\'exico 07000 D.F.}\\
 {\small\it$^2$Instituto de Ciencias Nucleares, Universidad Nacional
Aut\'onoma de M\'exico,}\\
{\small\it Apdo. Postal 70-543, M\'exico 04510 D.F.}\\
{\small\it$^3$ Departamento de F\'{\i}sica, Universidad Nacional de
La Plata, }\\ {\small\it c.c. 67 1900, La Plata, Argentina.}
}
\maketitle

\begin{abstract}
An exactly solvable model is introduced, which is equivalent to the exact
shell-model treatment of protons and neutrons in a single j-shell for
Fermi-type excitations.
Exact energies, quasiparticle numbers and double beta decay Fermi
amplitudes are computed and compared with the results of both the standard
quasiparticle
random phase approximation (QRPA) and the renormalized one (RQRPA), and
also with those corresponding to the hamiltonian in the
quasiparticle basis (qp). A zero excitation energy state is found in the
exact case, occuring at a value of the residual particle-particle
interaction at which the QRPA collapse. The RQRPA and the qp solutions do
not include this zero-energy eigenvalue in their spectra,
probably due to spurious correlations.
\end{abstract}

\noindent {\bf Resumen}
{\small
Se presenta un modelo exactamente soluble que es equivalente al modelo de
capas para protones y neutrones en una sola capa con momento angular j
para el caso de excitaciones tipo Fermi.
Se calculan los valores exactos de las energ\'{\i}as y de las amplitudes
del decaimiento beta doble tipo Fermi y se los compara con los resultados
de la aproximaci\'on de fases al azar para cuasipart\'{\i}culas (QRPA)
usual y con los de la renormalizada (RQRPA), y tambi\'en con los
correspondientes al hamiltoniano en la base de cuasipart\'{\i}culas (qp).
En el caso exacto se encuentra un estado con energ\'{\i}a de excitaci\'on
nula, que ocurre para el valor de la interacci\'on residual part\'{\i}cula
- part\'{\i}cula para el que la QRPA colapsa. La QRPA y las soluciones de
tipo qp no tienen este autoestado de energ\'{\i}a cero en su espectro,
probablemente debido a la presencia de estados espurios.
} 

\noindent
PACS number(s): 21.60.Jz, 21.60.Fw, 23.40.Hc

\vskip .5cm

\section{Introduction}

In order to improve the reliability of the Quasiparticle Random Phase
Approximation (QRPA) description of nuclear double beta decay transitions
much work has been done recently related with its extensions, in particular
with the Renormalized Quasiparticle Random Phase Approximation (RQRPA).

The matrix elements for ground state to ground state
two-neutrino double-beta decay transitions ( $\beta\beta_{2\nu}$) 
calculated in the QRPA are extremely sensitive to details of
the nuclear two-body interaction \cite{Vog86,Eng87,Civ87,Mut89}.

The inclusion of renormalized particle-particle correlations in the QRPA
matrix amounts to a drastic suppression of the
$ \beta\beta_{2\nu}$ -matrix elements. However,
for some critical values of the model parameters the QRPA
eigenvalue problem becomes complex. 
The most notorious example of this behavior, of the QRPA approach, is
the calculation of the $\beta\beta_{2\nu}$
decay of $^{100}Mo$ \cite{Vog86,Eng87,Civ91,Gri92}.

The renormalized version of the QRPA (RQRPA)\cite{Har64,Row68}, which
includes some corrections beyond the quasiboson approximation, has
been recently  reformulated \cite{Cat94} and applied to the
$\beta\beta_{2\nu}$ decay problem \cite{Toi95}. Contrary to the QRPA, the
RQRPA does not
collapse for any value of the residual two-body interaction.
Based on its properties, the RQRPA was presented as a cure
for the instabilities of the QRPA  and it was applied
to calculations of the $\beta\beta_{2\nu}$ decay of
$^{100}Mo$ \cite{Toi95}. Similar studies have been
performed in the framework of the RQRPA and with the
inclusion of proton-neutron pairing
correlations in symmetry breaking hamiltonians \cite{Sch96}.

In a recent paper \cite{Hir96a} we have shown that
the RQRPA violates the Ikeda sum rule and that this
violation is indeed present in many extensions of the QRPA.
The study was based on the schematic proton-neutron Lipkin model.

In the present paper we will review the general features of the exactly
solvable model and its comparison with the QRPA, the RQRPA and the
qp-hamiltonian \cite{Hir96a,Hir96b,Hir96c}. The presence of a zero
excitation energy state in the
spectrum corresponding to the exact solution of the model hamiltonian is
discussed. It will be shown that the RQRPA and the qp
solutions, do not display the same feature, most likely due to the
presence of spurious states caused by the mixing of orders, of
the relevant interaction terms, in the expansion procedure.

The structure of the paper is the following: the model and its solutions
are presented in
Section 2, the quasiparticle version of the hamiltonian, its linear
representation in terms of pairs of unlike (proton-neutron)
quasiparticle-pairs and
its properties are introduced in Section 3. The QRPA and RQRPA
treatments of the hamiltonian are
discussed in Section 4. The matrix elements of double-beta-decay
transitions, calculated in the framework of the different approximations
introduced in the previous sections, are given
in Section 5.
Conclusions are drawn in Section 6.

\section{The model}

The model hamiltonian, which includes a single  particle  term, a pairing
term for protons and neutrons and a schematic charge-dependent  residual
interaction with both particle-hole  and particle-particle channels, is
given by

\begin{equation}
H = e_p {\cal N}_p - G_p S^\dagger_p S_p +
 e_n {\cal N}_n - G_n S^\dagger_n S_n +
2 \chi \beta^- \cdot \beta^+
	  - 2 \kappa P^- \cdot P^+ , \label{hamex}
\end{equation}

\noindent with
\begin{equation}
\begin{array}{c}
{\cal N}_i = \sum\limits_{m_i} a^\dagger_{m_i} a_{m_i} ~,\hspace{1cm}
S^\dagger_i = \sum\limits_{m_i} a^\dagger_{m_i} a^\dagger_{\bar m_i}/2 ~,
\hspace{1cm}i=p,n\\
 \beta^- = \sum\limits_{m_p = m_n} a^{\dagger}_{m_p} a_{m_n} ~,
 \hspace{1cm}
P^- = \sum\limits_{m_p = -m_n}	a^{\dagger}_{m_p} a^{\dagger}_{\bar m_n} ~;
\end{array}
\end{equation}

\noindent
$ a^{\dagger}_p = a^\dagger_{j_p m_p}$ being the particle creation
 operator and
 $a^{\dagger}_{\bar p} = (-1)^{j_p -m_p} a^{\dagger}_{j_p -m_p}$ its time
reversal. The parameters $\chi$ and $\kappa$ play the role of the
renormalization factors $g_{ph}$ and $g_{pp}$ introduced in the
literature \cite{Vog86,Eng87,Civ87,Mut89}.

The relatively simple schematic force (1) can approximately describe the
correlations induced by a more realistic interaction.

In a single-one-shell limit, for the model space ($j_p = j_n = j$) and for
monopole ($J=0$) excitations the hamiltonian (1) can be solved exactly.
In spite of the fact that the solutions obtained in this restricted
model space cannot be related to actual nuclear states,
the excitation energies, single- and double-beta decay transition
amplitudes and ground state correlations depend on the particle-particle
strength parameter $ \kappa$ in the same way as they do in realistic
calculations with many single particle levels and with more realistic
interactions. 
We shall obtain the eigenstates of (1), by using different
approximations, in order to built-up a comprehensive view about the
validity of them and their predictive power.

The hamiltonian (\ref{hamex}) can be expressed in terms of the
generators of an SO(5) algebra \cite{Par65}.
The Hilbert space is constructed by using the eigenstates of the
particle-number operator ${\cal N} = {\cal N}_p + {\cal N}_n$ , the
isospin $\cal T$ and its projection ${\cal T}_z = ({\cal N}_n -
{\cal N}_p)/2$. The raising and lowering isospin operators are
defined as $\beta^\pm ={\cal T}^\pm$, where ${\cal T}^- |n\rangle =
|p\rangle$.
With them we can construct the isospin scalar ${\cal T}^2 $
and the second order SO(5) Casimir.

The hamiltonian (\ref{hamex}) is diagonal in the ${\cal N, T, T}_z$ basis
if
$G = 4 \kappa$. It can be
reduced to an isospin scalar if its parameters are selected as

\begin{equation}
e_p =e_n, \hspace{1cm}\chi=0,\hspace{1cm}G=4\kappa.
\end{equation}
If $4\kappa \neq G$ the hamiltonian (\ref{hamex}) will not be diagonal
in this basis. The
hamiltonian mixes states with different isospin $T$ while its
eigenstates still have definite $N$ and $T_z$. The dynamical breaking
of the
isospin symmetry is an essential aspect of the model which is
directly related to the nuclear structure mechanism
responsible for the suppression of the
matrix elements for double-beta-decay transitions.

\subsection{The diagonal case $G = 4\kappa$}

For the numerical examples we have selected
$N_n > N_p $ and a large value of $j$ to
simulate the realistic situation found in medium-  and heavy-mass
nuclei. To perform the calculations we have adopted the following
set of parameters:

\begin{equation}
\begin{array}{llll}
&j = 19/2, & N = 20, & 1 \le T_z \le 5,\\
&e_p = 0.69 MeV, & e_n = 0.0 MeV, \\
& G_p = G_n = 0.2 MeV, &\chi=0 ~\hbox{or}~ 0.025 MeV~,
&0 \le \kappa \le 0.1
\end{array}
\end{equation}

The dependence of the spectrum and transition matrix elements on
the parameters $\chi$ and $\kappa$ is analyzed in the following
paragraphs.

 {\bf Fig. 1}

The complete set of $0^+$ states, belonging to different isotopes,
is shown in Fig. 1a  for $G= 4\kappa$ and $\chi = 0.$,
as a function of the number of protons (Z).
The states are labeled by the isospin quantum numbers $(T,T_z)$.
Ground states are shown by thicker lines. As shown in the
figure the structure of the mass parabola is qualitatively
reproduced.

The upper insert, case a), shows the full spectrum
corresponding to $\chi = 0$. The lower one, case b),
shows the results corresponding to  $\chi = 0.025 MeV$.
Obviously the particle-hole channel of the residual
interaction stretches the spectra of all isotopes. 
It also increases the energy of the Isobaric Analog State (IAS).

Beta decay transitions of the Fermi  type, mediated by the action of the
operator $\beta^- =  t^-$, are allowed between states belonging to the same
isospin multiplet. The energy of each member of a given multiplet increases
linearly with $Z$.

In this example the  $0^+$ states belonging to each odd-odd-mass nuclei (
N-1, Z+1, A) are the IAS constructed from the $0^+$ states of the
even-even-mass nuclei with (N, Z, A) nucleons. Thus, Fermi transitions
between them are allowed.

Since the isospin of the ground state of each of the even-even-mass nuclei
differs, for different isotopes, Fermi-double-beta-decay transitions
connecting them are forbidden in this diagonal limit $G = 4 \kappa$.

\subsection{Exact solutions}

The hamiltonian (\ref{hamex}) has a ${\cal T}=2$ tensorial component
which mixes states with different isospin, while particle number
and isospin projection remain as good quantum numbers.	The
diagonalization of (1) is performed in the basis of states belonging to 
the SO(5) irrep $(\Omega,0)$ \cite{Hir96c}.

\begin{equation}
|{\cal N T}_z \alpha \rangle = \sum\limits_{\cal T} {\cal
C}^\alpha_{{\cal N T T}_z} |{\cal N T T}_z \rangle
\end{equation}

{\bf Fig. 2}

\bigskip

The energies of the ground-state ($0_{g.s}^+$) and of the first-excited
state ($0_1^+$), as a function of the ratio $4 \kappa /G$
for ${\cal N}_n = 12, ~{\cal N}_p = 8$ and $\chi=0$ are shown if Fig. 2.

The most characteristic feature of the results is the barely
avoided crossing of levels, due to the repulsive
nature of the effective residual interaction between them. 
In the neighbourhood of the value $4 \kappa /G \approx 1 $ 
a major structural change in the wave functions will develop. 

\bigskip
{\bf Fig. 3 }

\bigskip

The full-thin line of Fig. 3a (3b) represents the excitation
energy	$E_{exc}$ of the lowest $0^+$ state belonging to the
double-odd-mass nucleus (${\cal N}_n=13, {\cal N}_p= 7$)
with respect to the parent even-even-mass nucleus
(${\cal N}_n=14, {\cal N}_p=6$) , as a function of the ratio
 $4 \kappa / G$ for  $\chi=0$  ($0.025$).
It is clear that when $4 \kappa / G \approx 1.3  $ attractive
proton-neutron correlation dominates over  proton-proton and
neutron-neutron pairing correlations and the  excitation energy goes to
zero.

The vanishing of the energy of the first excited state,
and the subsequent inversion of levels (or negative excitation energies)
would indicate that the double-odd nucleus becomes more bound
than their even-even neighbours, contradicting the main
evidence for the dominance of like-nucleons pairing
in medium- and heavy-mass nuclei. 
It would also completely suppress
the double beta decay because the single beta decay from each
"side" of the double-odd nucleus would be allowed.

These result simply emphasizes the fact that the
hamiltonian (\ref{hamex}) will not be the adequate one when
attractive proton-neutron interactions are too large.
In a realistic situation, obviously, the true hamiltonian
includes other degrees of freedom, like
quadrupole-quadrupole interactions,  and permanent deformations
of the single-particle mean-field can also be present.
These additional degrees of freedom will prevent the complete
crossing of levels which, of course, is not observed. However,
in many cases the experimentally observed energy-shift of double-odd-mass
nuclei, respect to their double-even-mass neighbours
is very small. This finding reinforces the notion of an underlying
dynamical-symmetry-restoration-effect.

\section{The hamiltonian in the quasiparticle (qp) basis}

By performing the transformation of the particle creation and
annihilation operators of the hamiltonian (1) to the quasiparticle
representation \cite{Row70} we have obtained the qp-hamiltonian
\cite{Hir96b,Hir96c}.

The linearized version of the qp-hamiltonian is obtained by neglecting
the scattering terms.
The solutions of this truncated hamiltonian have been discussed
in \cite{Hir96a}.

Finding the eigenvalues and eigenvectors  of the qp-hamiltonian
requires the use of the same algebraic techniques involved in
solving the original hamiltonian. However, the	complexity of the
problem increases severely, due to the fact that neither the
quasiparticle number or the quasiparticle isospin projection
(or equivalently the number of proton and neutron quasiparticles)
are good quantum numbers. It implies that the dimension of the basis
will increase by two orders of magnitude \cite{Hir96c}.

Particle number is not a good quantum number, obviously, because
it is broken spontaneously by the Bogolyubov
transformation. Thus, zero-quasiparticle states belonging
the even-even-mass nucleus have good {\em average} number of protons and
neutrons while states with a non-vanishing number
of quasiparticles show
strong fluctuations in the particle number. Fluctuations in the
particle number can induce, naturally,
important effects on the observables. Moreover, the
admixture of several quasiparticle-configurations in a given state,
induced by residual particle-particle interactions, can also strongly
influence the behavior of the	observables. An example of this
effect is given in \cite{Hir96a}, concerning the violation of the
Ikeda Sum Rule produced by
large values of the particle-particle strength $\kappa$.

The spectrum of the qp-hamiltonian for odd-odd nuclei is shown in
Figure 3.
The curves shown by small-dotted-lines, in Figs.  3.a), 3.b),
display the dependence of the
excitation energy for the  qp-hamiltonian upon
the ratio $\kappa /G$. The energies in  this qp-approximation
closely follow the exact ones
up to the point where the last become negatives ($4 \kappa /G \approx 1.4~
-~1.8$ in the different cases). From this point on
they vanish, rather than taking negative values, instead.
The excitation energies for the linearized hamiltonian $H_{22}+H_{04}$
are shown as thick lines in these figures. We can see that the
linearized hamiltonian is able to reproduce qualitatively the
behavior of the full-qp one, but in general it overestimates the
values of the excitation energies.

As it is mentioned above, the results shown in 
Figures 3 have been obtained
both with the complete qp-hamiltonian and
 with the truncated hamiltonian which includes
only the product of pair-creation and annihilation-
operators. In \cite{Hir96a} the relevance of the scattering terms in
the qp-hamiltonian was pointed out. From the present results
it can be seen that the inclusion of these terms is indeed
important if one looks after a better description
of the qp-excitation energies, up to
the point where the exact excitation energies become negative.
For
larger values of $\kappa$ even the eigenstates of the complete
hamiltonian fail to describe negative excitation energies. This is
a clear indication that other
effects can play an important role, like, i.e; effects associated
to the appearance of spurious
states. This can be quantitatively illustrated by the following.
There are four exact eigenstates for
$j=19/2, ~{\cal N}_n = 13, ~{\cal N}_p =7$, as can be seen in Fig. 2.a),
while the spectrum of the qp-hamiltonian has 220 eigenstates.
It is
well known that states with ${\cal N}_n = 14 \pm N_n, ~{\cal N}_p = 6
\pm N_p$, where $N_p$ and $N_n$ are the number of quasiparticle protons
and neutrons,
respectively, are mixed with the two p-n quasiparticle state in the
odd-odd nucleus and provide a large number of states belonging to
other nuclei. When $4 \kappa /G \ll 1$ the spurious states remain
largely un-mixed with the lower energy two-qp state. But when
$4\kappa/G \approx 1$ the mixing becomes important.
This fact upgrades the relevance of particle-number violation
effects in dealing with this case.

The full qp-treatment represents the best possible extension of
the quasiboson
approximation, without performing a
particle-number projection, in a single-j shell. It goes beyond any
second extended RPA \cite{Mar90} and it includes explicitly all
number of proton-
and neutron-quasiparticles ($N_p$ and $N_n$) in the eigenstates.

To analize the effects associated to the number of
quasiparticles in the ground state of double-even nuclei,
and particularly the effects associated to the number of
quasiprotons,
we have calculated the average number of quasi-protons
\cite{Hir96a,Hir96c}.

{\bf Fig 4}

In Figs. 4.a) and 4.b) the average number of proton-quasiparticles in the
ground state of the
even-even nucleus with ${\cal N}_p=6,~{\cal N}_n=14$ is shown as a
function of $4 \kappa/G$, for $j=19/2,
~\chi=0 $ and $0.04$.  The
dashed-lines represent the results corresponding to the full
qp-hamiltonian case while the large dots refer to the linearized $H_{22}
+ H_{04}$ version of it.
The difference between both approximations is evident. Using the
linearized hamiltonian
the states are composed only by proton-neutron-quasiparticle pairs
\cite{Hir96a}, while the presence
of the scattering terms introduces also the like quasiparticle
pairs. The presence of these pairs, which for
$4\kappa /G \approx 1$ play a crucial role, increases
notably the number of
quasiparticles and yields excitation energies closer to the exact ones.

The average quasiparticle number shows a saturation in the full-qp
case for $4\kappa /G \approx 1.8$. 
At this value of the residual pn-interaction
the ground state is far-away for the qp vacuum, and has a structure
which can be described as {\em full quasiparticle shell}. Notice that 
at this point the
exact and full-qp excitation energies depart for each other. A state
with four proton
and four neutron quasiparticles has very large number-fluctuations.
Spurious states
become strongly mixed with physical states. In this
way the resulting excitation
energies average to zero, a limit which
differs from the exact value, which is negative.

\section{QRPA and RQRPA}

The QRPA hamiltonian $H_{QRPA}$ can be obtained from the linearized
version of the qp-hamiltonian, by keeping only the bilinear-terms
in the pair-creation and pair-annihilation operators.
The pair-creation and pair-annihilation operators are defined
by coupled pairs of fermions. The commutation
relations between these pseudo-boson operators include number-like
quasiparticle operators in addition to unity. By
taking the limit $(2j+1) \rightarrow \infty$ \cite{Hir96a}
these extra terms vanish and the commutation relations between
pairs of fermions can be treated like exact commutation relations
between bosons. This is the well-known quasi-boson approximation.

The QRPA states are generated by the action of the one-phonon operator
on the correlated QRPA vacuum. If ground state correlations are too strong 
the first eigenvalue becomes purely imaginary.
For this limit the backward-going amplitudes of the QRPA
phonon-operator become dominant, thus invalidating the
underlying assumption about the smallness of the
quasi-boson vacuum-amplitudes.
The QRPA excitation energies, obtained with the above introduced
hamiltonian are shown in Figs. 3a, 3b.
It can be seen that in the cases displayed in these figures
the collapse of the QRPA values occurs near
the point where the exact excitation energies become negative.
This is a very important result because it means that
that the QRPA description of
the dynamics given by the hamiltonian (\ref{hamex}) is able to reproduce
exact results. At this point one can naturally ask the obvious
question about the nature of the mechanism
which produces such a collapse. The fact that the QRPA approximation
is sensitive to it, together with the fact that the
same behavior is shown by the exact solution, reinforces the idea
about the onset of correlations which terminate the regime of
validity of the pair-dominant picture.
In order to identify such correlations we have calculated the
expectation value of the number of quasi-fermions and bosons
on the	QRPA ground state \cite{Hir96a,Hir96c}.
Figs. 4.a), 4.b)  show the results corresponding to these
occupation numbers. The QRPA results extend up to
the value $4 \kappa
/ G \approx 1$, where the QRPA collapses. The sudden increase of the
average quasiparticle
number near the collapse of the QRPA is a clear evidence
about a change in the structure of
the QRPA ground state.

In the renormalized QRPA the structure of the ground state is
included explicitly \cite{Row68}.
The quasi-boson approximation is not enforced explicitly. The
renormalization procedure consists of
retaining approximately the number of quasiparticle-like-terms of the
commutators keeping them as a parameter to be determined.
It produces the
reduction of the residual interaction  which is needed to avoid
the collapse of the QRPA equations \cite{Toi95}.
Due to this fact the RQRPA energy $E_{RQRPA}$ is always real. Its
value can be obtained by solving simultaneously a set of non-linear
equations \cite{Toi95}.

 RQRPA	excitation energies are shown in Figs. 3a and 3b.
The main finding of the present calculations is that the exact
excitation energies are closer to the QRPA energies, rather than to
the renormalized ones.
In exact calculations including the spin degrees of freedom a phase
transition was found at the point where the QRPA collapses \cite{Eng96},
thus reinforcing the present results.

The average number of quasiparticles in the RQRPA vacuum 
is shown in Fig. 4.a), 4.b).
It is fairly obvious, from these results, that the RQRPA
ground state correlations double in all the cases those of the complete
solutions
of the linearized hamiltonian. This is clearly an overestimation, and it
is probably one of the most
notorious difficulties confronting the use of the RQRPA.

It allows too
much ground state
correlations, and with them the particle number fluctuations are
introducing spurious
states which can dominate the low energy structure for
large values of $\kappa$.

Near "collapse" the average number of quasiparticles given by the QRPA
and the RQRPA are comparable.
For the case of the QRPA the increase of the ground-state-correlations
is determined by the change in the sign of the backward-going
matrix relative to the forward-going one near collapse. From
there on the QRPA
cannot produce any physically acceptable result since one of
the underlying conditions of the approximation, i.e: the
positive definite character of either linear combination of the
forward- and backward-going blocks of the QRPA matrix will not be
fulfilled. This collapse is prevented in the RQRPA, by the
use of the renormalization of the matrix elements, but the drawback
of the approximation is the
contribution coming from spurious states, which ought to be removed.
By going beyond the leading order QRPA approximation, more
terms have to be added to the diagrams which represent the
transition amplitudes. 
It has been done for a pure seniority model in \cite{Duk96}.

\section{Double beta decay}

 In this section we shall briefly discuss some of the consequences
of the previously presented approaches on the calculation
of nuclear double-beta decay observables. In the following
we shall focus our attention on the two-neutrino mode of the
nuclear double-beta decay, since the matrix elements governing
this decay mode are more sensitive to nuclear structure effects
than neutrinoless mode. As said in the introduction we shall
consider only double-Fermi transitions.
The nuclear matrix elements of the two-neutrino
double-beta-decay  $M_{2\nu}$ are discussed in \cite{Hir96a,Hir96b,Hir96c}

The results for the matrix elements $M_{2\nu}$, obtained with the
exact wave functions are
shown, as a function of the ratio $4 \kappa/ G$,
in Figs. 5.a) and  5.b). These results have been obtained
with the following set of parameters:
$j=19/2, ({\cal N}_p=6,~{\cal N}_n=14)\rightarrow ({\cal
N}_p=8,~{\cal N}_n=12)$ and $\chi = 0 $ and $0.025 MeV$.

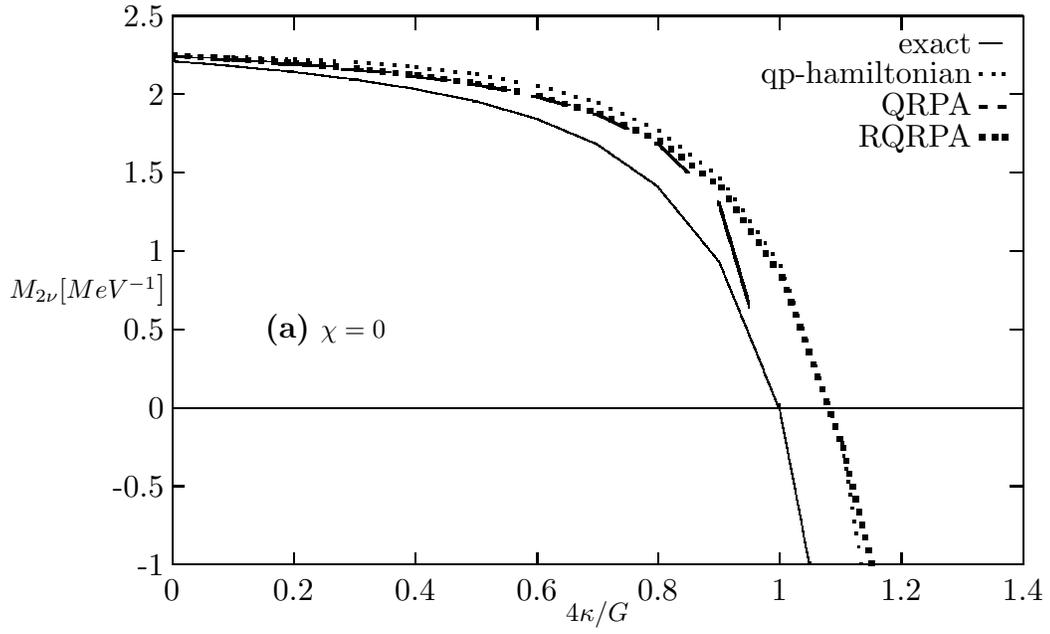
{\bf Fig. 5}

The exact value of
 $M_{2\nu}$ vanishes at the point $4 \kappa / G = 1$.
As mentioned above, this cancellation
appears in the model due to the fact that for this value of $\kappa$ the
isospin-symmetry is recovered and the ground states of the initial
and final nuclei belong to different isospin multiplets, as it can be seen
also from the results shown in Fig. 1.

A similar mechanism, in the context of a solvable model
possessing a SO(8) algebra including spin
and isospin degrees of freedom
was used a decade ago to shown that the
cancellation of the
$M_{2\nu}$ matrix elements for certain values of the particle-particle
residual interaction was
not an artifact of the QRPA description \cite{Eng87}.

The results corresponding to the matrix elements $M_{2\nu}$,
calculated with the different approximations discussed in the
text are also shown in Fig. 5, as a function of the coupling
constant $\kappa$.
The values  of $M_{2\nu}$ are very similar to those
found in realistic
calculations \cite{Vog86,Civ87,Mut89,Toi95}, including its strong
suppression for values of the coupling constant $\kappa$
near the value which produces the collapse of the QRPA description.
Distinctively, the RQRPA results extends to values
of $\kappa$ passing the "critical" value. However, the validity of this
result can be questioned because, as we have shown above,
the RQRPA missed the vanishing of the excitation energy.
The $M_{2\nu}$ matrix elements, evaluated
with the complete qp-hamiltonian, is quite similar to that of
the RQRPA up to point where it vanishes
From this point-on the results of both the full-qp
and the RQRPA approximations  are different. Both matrix elements
change their sign at a
value of $\kappa$ which is larger than the one corresponding
to the change of the sign of the matrix elements calculated with the
exact wave function. The fact that the RQRPA results and the ones of
the qp-approximation are similar, although these models differ
drastically in the correlations which they actually include, suggest
that a kind of balance is established between terms which are
responsible for ground state correlations and those which produce
the breaking of coherence in the wave functions. Obviously this
mechanism must be related to the presence of scattering terms
in the commutators as well as in the hamiltonian.

\section{Conclusions}

An exactly solvable model for the description of single- and double-beta-
decay-processes of the Fermi-type was introduced.
The model is equivalent to a complete shell model treatment in a
single-j shell for the adopted hamiltonian.
It reproduces the main features of the results obtained in
realistic calculations, with
many shell and effective
residual interaction, like those used in the literature to describe the
microscopic structure of the nuclei involved in double
beta decay processes.

We have constructed the exact spectrum of the hamiltonian
and discussed its properties. The results concerning the
energy of the states belonging to the exact solution of the
model show that, in spite of its very schematic structure,
the hamiltonian is able to qualitative reproduce the nuclear mass
parabola. The sequence of levels of the exact
solution shows
that the ground-state and the first-excited state,
of the spectrum of double-even nuclei, approach
a band-crossing situation for a critical value of the
strength associated to attractive particle-particle interactions.
At the crossing these states interchange their quantum numbers.
This behavior is connected with the
description of "shape" transitions in similar theories, where the
order parameter is clearly associated with multipole deformations
in r-space. In the present model the "deformation" mechanism is
related with the breaking of the isospin symmetry and the space-rotation
correspond to a rotation in isospin-space which preserves the
third-component of the isospin.

We have compared the exact values of the excitation energy and of the
double-beta-decay matrix elements, for double-Fermi transitions,
with those obtained by using the solutions of the
approximate qp-hamiltonian, its linearized version and both the QRPA and
RQRPA ones.

It was shown that the collapse of the QRPA correlates with
the presence of
an exact-eigenvalue at zero energy. The structure of the RQRPA solutions
has been
discussed and it was found that though finite they are not free from
spurious contributions. The role of scattering-terms was discussed and
they were shown to be relevant in getting excitation energies closer
to the exact values. However they are not enough
to generate the correlations which are needed to produce
the band-crossing, or negative excitation energies, as it was found
in the exact solution  for large values of the coupling constant
$\kappa$.

In order to correlate the break-up of the QRPA approximation
with the onset of strong fluctuations in the particle number
we have calculated the average number of quasiparticles in the
different approximations discussed in the text.

It was shown that the solutions of the complete qp-hamiltonian
display a strong change in
the structure of the ground state when the particle-particle strength
increases.
The qp-content of the ground state varies
from a nearly zero-value to an almost full qp-occupancy.
The particle
number fluctuations
associated with states with a large number of quasiparticles were
mentioned as a possible source of spurious states.

Double beta decay amplitudes were evaluated in
the different formalisms. Their
similitudes and differences were pointed out.

As a conclusion the need of additional
work, to clarify the meaning of the different approximations
posed by the RQRPA, was pointed out.

\section{Acknowledgments}

Partial support of the CONACyT of Mexico and the CONICET of
Argentina is acknowledged. O.C. acknowledges a grant
of the J. S. Guggenheim Memorial Foundation.

\newpage

\centerline{\bf Figure Captions}

\bigskip
Figure 1.a (1,b):  $0^+$ states of different isotopes are shown
for $j=19/2, ~4\kappa /G =1$
and $\chi = 0. ~(0.025) MeV$.
in an energy vs. $Z$ plot.
States are labeled by $(T,T_z)$.
The lowest energy state of each nucleus is
shown by a thick-line.

\bigskip
Figure 2: Energy of the ground state $0_{gs}^+$ (full line)
and first excited
state $0_1^+$ (dotted line), as a function of the ratio
$4 \kappa /G$, for $j=19/2,~{\cal N}_n = 12, ~{\cal N}_p = 8$.

\bigskip
Figure 3.a (3.b): Excitation energy  $E_{exc}$ of the lowest $0^+$
state in the odd-odd intermediate nucleus (${\cal N}_n=13, {\cal N}_p= 7$) 
with respect to the parent even-even nucleus
(${\cal N}_n=14, {\cal N}_p=6$) against $4 \kappa / G$, for
$j=19/2,~\chi=0$ ($0.025$).
Exact results are shown as thin-full-lines while those of the
qp-hamiltonian are shown as small-dotted-lines. Results
corresponding to the linearized qp-hamiltonian
are shown as full-thick-lines and the results obtained with
the QRPA and RQRPA methods as large-dotted- and dashed-lines,
respectively.

\bigskip
Figure 4.a (4.b):  Average number of proton-quasiparticles
 in the ground state of the even-even nucleus with ${\cal N}_p=6,~{\cal
N}_n=14$  as  function of $4 \kappa/G$, for $j=19/2, ~\chi=0 ~(0.025)
MeV$.
Results corresponding to the qp-hamiltonian are shown as
dashed-lines. The ones corresponding to the linearized
 qp-hamiltonian are shown  as large-dotted lines and those of the
QRPA and RQRPA methods as full-lines and small-dotted-lines,
respectively.

\bigskip
Figure. 5a (5b): Matrix elements  $M_{2\nu}$, for the
double-Fermi two-neutrino double-beta decay mode, as
functions of the ratio $4 \kappa/ G$
for $j=19/2, ({\cal N}_p=6,~{\cal N}_n=14)\rightarrow ({\cal 
N}_p=8,~{\cal N}_n=12)$ and $\chi = 0 ~(0.025) MeV$
 Exact results are indicated by thin-full-lines.
The results obtained with the
qp-hamiltonian are shown as small-dotted-lines and the results of the
QRPA and RQRPA methods as dashed-lines and large-dotted-lines,
respectively.

\newpage
~
\vskip -3.cm
\centerline{Figure 1}

\vskip 1cm

\setlength{\unitlength}{0.240900pt}
\ifx\plotpoint\undefined\newsavebox{\plotpoint}\fi
\sbox{\plotpoint}{\rule[-0.200pt]{0.400pt}{0.400pt}}%




\bigskip
\begin{thebibliography}{Hir94b}
\bibitem{Vog86} P. Vogel and M. R. Zirnbauer, Phys. Rev. Lett. {\bf 57}
(1986) 3148.
\bibitem{Eng87} J. Engel, P. Vogel and M. R. Zirnbauer, Phys. Rev {\bf
C37} (1988) 731.
\bibitem{Civ87} O. Civitarese, A. Faessler and T. Tomoda, Phys.
Lett. {\bf B194} (1987) 11.
\bibitem{Mut89} K. Muto, E. Bender and H. V. Klapdor, Z. Phys. {\bf
A334} (1989)
\bibitem{Civ91}  O. Civitarese, A. Faessler, J. Suhonen and X. R. Wu,
Nucl. Phys. {\bf A524} (1991) 404; J. Phys. G: Nucl. Part. Phys. {\bf 17}
(1991) 943.
\bibitem{Gri92} A. Griffiths and P. Vogel,  Phys. Rev. {\bf C 46}
(1992) 181.
\bibitem{Har64} K. Hara, Progr. Theor. Phys. {\bf 32} (1964) 88;
K. Ikeda, T. Udagawa and Y. Yamaura, Prog. Theo. Phys. {\bf 33}  (1965) 22.
\bibitem{Row68} D. J. Rowe, Phys. Rev. {\bf 175} (1968) 1283; Rev. Mod.
Phys. {\bf 40} (1968) 153; J. C. Parick and D. J. Rowe, Phys. Rev. {\bf
175} (1968) 1293; D. J. Rowe, Nucl. Phys. {\bf A 107} (1968) 99.
\bibitem{Cat94} F. Catara, N. Dinh Dang, M. Sambataro, Nucl. Phys. {\bf
A 579} (1994) 1.
\bibitem{Toi95} J. Toivanen and J. Suhonen, Phys. Rev. Lett. {\bf 75}
(1995) 410.
\bibitem{Sch96} J. Schwieger, F.Simkovic and Amand Faessler, Nucl. Phys.
{\bf A 600} (1996) 179.
\bibitem{Hir96a} J. G. Hirsch,	P. O. Hess and O. Civitarese, Phys.
Rev. {\bf C 54} (1996) 1976.
\bibitem{Hir96b} J. G. Hirsch,	P. O. Hess and O. Civitarese, Phys. Lett.
{\bf B} in press.
\bibitem{Hir96c} J. G. Hirsch,	P. O. Hess and O. Civitarese, Phys.
Rev. {\bf C} submitted. 
\bibitem{Par65} J. C. Parikh, Nucl. Phys. {\bf 63} (1965) 214; K. T.
Hecht, Nucl. Phys. {\bf 63} (1965) 177; A. Klein and E. R. Marshalek,
Rev. Mod. Phys. {\bf 63} (1991) 375.
\bibitem{Row70} D. J. Rowe, {\em Nuclear Collective Motion},
Methuen and Co. Ltd., London 1970.
\bibitem{Mar90} A. Mariano, J. Hirsch and F. Krmpoti\'c, Nucl. Phys.
{\bf A 518} (1990) 523 and references therein.
\bibitem{Eng96} J. Engel, S. Pittel, M. Stoitsov, P. Vogel and J. Dukelsky,
Los Alamos Preprint nucl-th/9610045.
\bibitem{Duk96} J. Dukelsky, P. Schuck, Phys. Lett. {\bf B 387} (1996) 233.
\end{thebibliography}
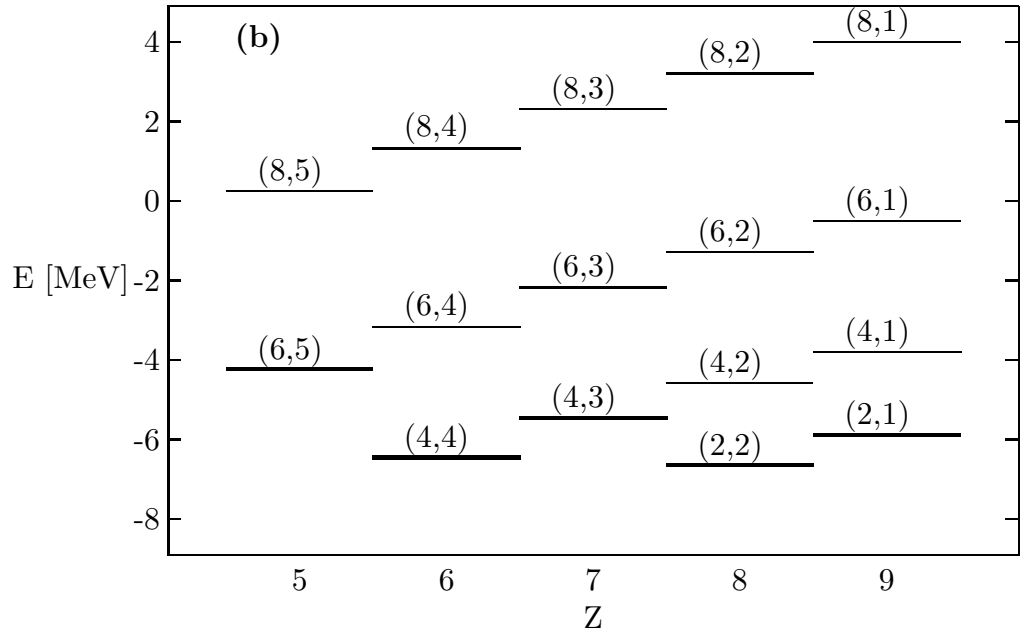
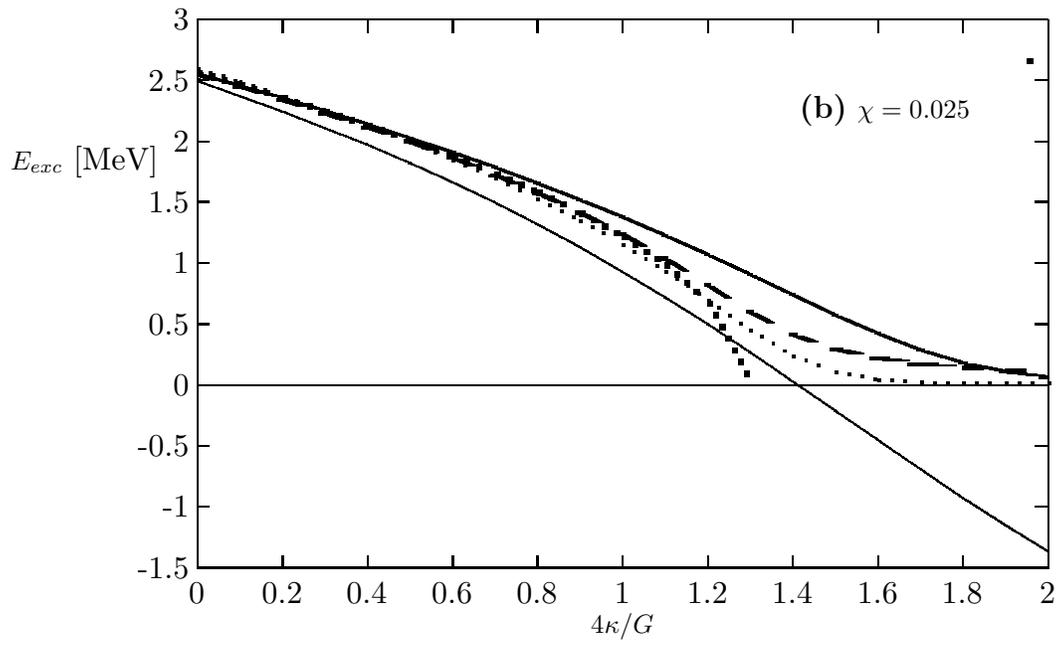
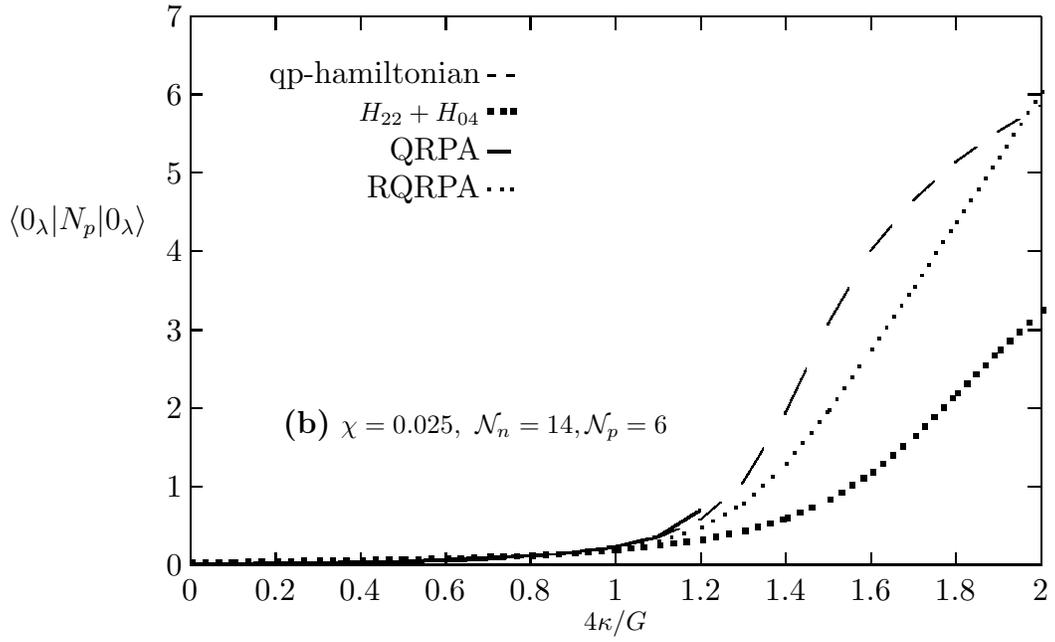
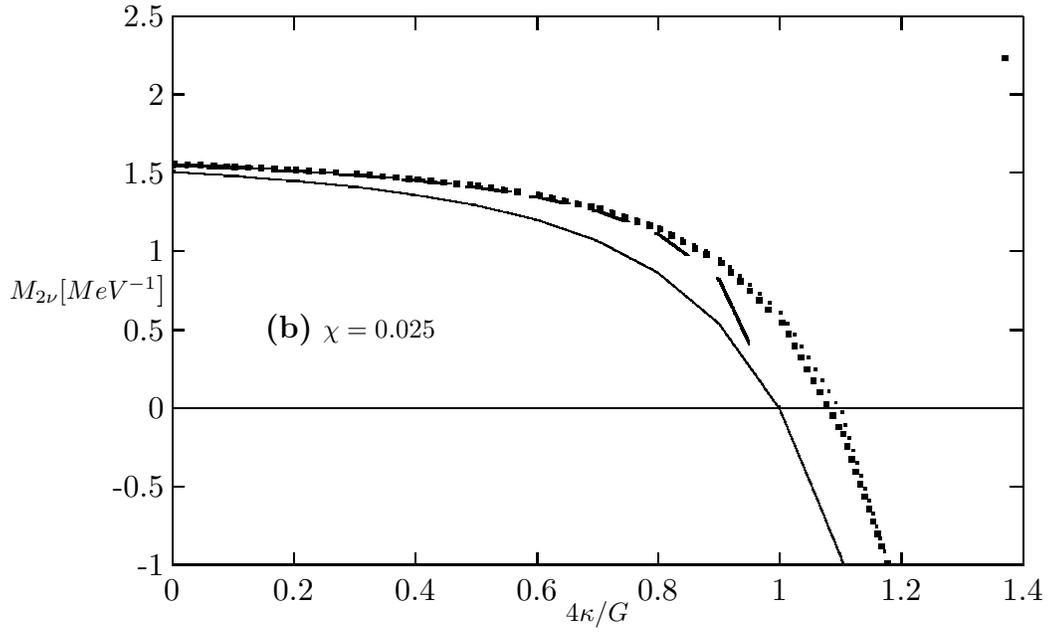
\end{document}